\documentclass[11pt]{IEEEtran}
\usepackage{amssymb}
\usepackage{hyperref}
\usepackage{amsmath}
\usepackage{braket}
\usepackage{cuted}
\usepackage{graphicx}
\usepackage{algorithm}
\usepackage[noend]{algpseudocode}
\usepackage{soul}

\newcommand{\qed}{$\Box$}
\newenvironment{widetext}
        {%
            \begin{strip}
            \par 
        }%
        {%
            \par
            \end{strip}
        }

\begin{document}

\title{Signed quantum weight enumerators\\ characterize qubit magic state distillation }
\author{Patrick Rall, \textit{Quantum Information Center, University of Texas at Austin}\\ \today\\ \vspace{-10mm}}
\maketitle

\begin{abstract}
    Many proposals for fault-tolerant quantum computation require injection of `magic states' to achieve a universal set of operations. Some qubit states are above a threshold fidelity, allowing them to be converted into magic states via `magic state distillation', a process based on stabilizer codes from quantum error correction. 
    
    We define quantum weight enumerators that take into account the sign of the stabilizer operators. These enumerators completely describe the magic state distillation behavior when distilling $\ket{T}$-type magic states. While it is straightforward to calculate them directly by counting exponentially many operator weights, it is also an NP-hard problem to compute them in general. This suggests that finding a family of distillation schemes with desired threshold properties is at least as hard as finding the weight distributions of a family of classical codes. 

    Additionally, we develop search algorithms fast enough to analyze all useful 5 qubit codes and some 7 qubit codes, finding no codes that surpass the best known threshold.
\end{abstract}

Protection of quantum states via quantum error correction codes is crucial for achieving experimental quantum computation \cite{preskill}. However, implementing a full set of universal gates can be challenging inside an error correction code \cite{ZCC07}. In 2004, Bravyi and Kitaev showed that Clifford operations, a non-universal gate set, can be extended to universality via injection of certain `magic states'. Furthermore, they showed that magic states could be distilled from several copies of lower-fidelity mixed resource states \cite{bk04}. This is known as Quantum Computing via State Injection (QCSI). More than a decade later, we still do not have a complete classification of mixed single-qubit states in terms of usefulness in QCSI, or a definitive method to distill an arbitrary useful state.

A discrete Wigner function defined on prime-dimensional qudit states is a helpful tool for this classification. In the case of odd primes negativity of the Wigner function is a necessary condition for usefulness of QCSI resource states \cite{vcge12}. In the case of qubits the situation is more complicated, and is a subject of active study. Wigner function negativity is no longer necessary for qubit resource states \cite{rbdobv15}, and must be replaced with a classification in terms of non-contextual hidden variable theories  \cite{rbdobv16}. For odd-dimensional qudits these two notions coincide \cite{dovbr16}, so classification is easier.

Resource states are distilled into magic states using stabilizer codes \cite{cb-struct}. Several copies of the resource state $\rho$ are projected onto the code space of a stabilizer code, and an output state $\rho'$ is decoded. Codes vary in terms of the rate at which they improve the fidelity of resource states, and often several iterations are necessary to obtain a high-fidelity magic state. Codes can also worsen the fidelity of resource states, and have a threshold input fidelity that is required for fidelity improvement. Classification of resource states can be achieved by finding stabilizer codes with low-enough threshold fidelity \cite{R06}, because they give an explicit scheme for constructing magic states from low-fidelity resource states.

Our current picture of qubit resource states stems primarily from such explicit code constructions found in 2004 and 2006. Current works are attempting to classify qubit states using non-contextual hidden variable theories, which worked very well for odd prime dimensions \cite{hwve14, rbdobv16}. But even if a classification theory is available, it does not give an obvious method for code construction and searching over codes and states may still be needed \cite{dh15, acb12}. In the end, an explicit distillation scheme will be necessary for use in experiment. Furthermore, we cannot rule out the existence of resource states that according to hidden variable theories should be useful, but for some reason still cannot be distilled by Clifford operations. Such resource states, if they exist, would be exciting because they might give a model of computation slightly weaker than BQP.

Thus, understanding the distillation thresholds of stabilizer codes remains pivotal for QCSI research. The study presented in this paper is restricted to qubit systems, and codes with a single-qubit code space. After reviewing preliminaries in section~I, section~II derives the main result: \textit{the distillation behavior of a stabilizer code is fully characterized by a signed version of its weight enumerators}. In section~III we briefly compare these to previous studies of quantum weight enumerators \cite{sl96, r96} from the late 1990s, and show that computing them in general is NP-hard. In section~IV we show how to extract interesting distillation properties from the weight enumerators, and define a `distillation polynomial' summarizing this information. In section~V we analyze a class of codes with a particular transversal gate.

 \section{Preliminaries}

For every $[[n, 1]]$ stabilizer code $S$ there is a quantum circuit $C_\text{decode}$ composed entirely of Clifford gates that extracts the logical qubit of the code to the first physical qubit. We use this to define a `recovery' superoperator $\mathcal{R}$ from operators in the physical Hilbert space to operators in the code space. If $\Pi_S$ is a projector onto the code space then:

\begin{equation}
    \mathcal{R}(A) =  \text{Tr}_{\text{2,..,}n}(C_\text{decode} \Pi_S A \Pi_S C_\text{decode}^\dagger)
\end{equation}

\noindent where $\text{Tr}_{\text{2,...,}n}$ is a trace over all qubits but the first. $\mathcal{R}$ can act on both density matrices and unitary operators. 

All $n$-to-1 magic state distillation procedures $\rho \to \rho'$ with a single output qubit can be brought into the form $\rho \to \rho' = \mathcal{R}(\rho^{\otimes n})/\text{Tr}(\mathcal{R}(\rho^{\otimes n}))$ for some code $S$ without reducing distillation quality \cite{cb-struct} . Therefore understanding $\mathcal{R}$ is the key to understanding magic state distillation thresholds.

Let $\mathcal{P}_n$ be the $n$-qubit Pauli group. For $P \in \mathcal{P}_n$, The weight of a Pauli operator $\text{wt}(P)$ is the number of qubits it acts on non-trivially, i.e. not with the identity operator. We denote phase of an operator with $\lambda(P)\in \{\pm 1, \pm i\}$. An $[[n,1]]$ stabilizer code $S$ is a \textit{list} of $n-1$ commuting Pauli operators that generate an abelian subgroup $G \subset \mathcal{P}_n$. The code space is the common $+1$ eigenspace of these operators. For this to be well defined all $P\in G$ must have real eigenvalues, and thus real $\lambda(P)$. The product of two commuting Pauli operators with real phase also has real phase, and Clifford operations preserve realness of phase.

Let us inspect how $\mathcal{R}$ acts on $\mathcal{P}_n$. The code space projector can be written like this: $\Pi_S = \prod_{Q \in S} \frac{1+Q}{2}$.
If a Pauli operator $P \in \mathcal{P}_n$ anticommutes with some $Q \in S$ then $(1+Q)P(1+Q) = 0$, so $\mathcal{R}(P) = 0$. Preservation of the code space of a stabilizer code means to commute with the stabilizer $G$. We have $\mathcal{R}(P) \neq 0$ if and only if $P \in G^\perp$ where $G^\perp \subset \mathcal{P}_n$ is the normalizer of $G$. 

We know that if $P$ is a Pauli operator then either $\mathcal{R}(P)$ is also a Pauli operator or $\mathcal{R}(P) = 0$. The normalizer $G^\perp$ contains all Pauli operators that correspond to valid logical operations. Therefore for $P \in  G^\perp$ we have $\mathcal{R}(P) \in \mathcal{P}_1$, and for $P \not\in G^\perp$ we have $\mathcal{R}(P) = 0$. Since $\mathcal{R}$ preserves group structure, we see that restricting $\mathcal{R}$ to $G^\perp$ gives a homomorphism from $G^\perp \to \mathcal{P}_1$.

Since the code space consists of eigenstates of $G$, operators $P \in G$ leave encoded states unchanged. Thus $\mathcal{R}(P) = I$ if $P \in G$, i.e. $G$ is the kernel of $\mathcal{R}$. Similarly if $P \in -G$ then $\mathcal{R}(G) = -I$. The rest of $G^\perp$ must be mapped to $\pm X, \pm Y, \pm Z$. If we have an operator $\bar X$ where $\mathcal{R}(\bar X) = X$, then for any $Q \in G$ we have $\mathcal{R}(\bar XQ) = X$. Therefore the set of elements mapped to $X$ is the coset $\bar XG$. Similarly, the rest of $G^\perp$ can be divided into cosets, where each coset has a single real-phased element of $\mathcal{P}_1$ as its image.

In summary, $\mathcal{R}$ sends elements of $\mathcal{P}_n$ to $0$ unless they are in $G^\perp$. $\mathcal{R}$ is a homomorphism from $G^\perp$ to $\mathcal{P}_1$ with $G$ as its kernel. Each coset of $G$ in $G^\perp$ is mapped to one of $\pm I, \pm X, \pm Y, \pm Z$. This classification will be very important in the following derivation.

 \section{Derivation of Signed Quantum Weight Enumerators}

Now we explore how $\mathcal{R}$ acts on density matrices interesting for magic state distillation. Employing the Bloch sphere formalism, an arbitrary single-qubit state can be written using a three-component vector $\vec a$ with magnitude $\leq 1$:
$$\rho(\vec a) = \frac{1}{2} \left( I + a_XX + a_YY  + a_ZZ\right).$$

The $\ket{H}$ and $\ket{T}$ magic states are defined by the Bloch vectors $\vec a = (1,1,0)/\sqrt{2}$ and $\vec a = (1,1,1)/\sqrt{3}$ respectively, as in \cite{bk04}. $\ket{H}$ is an eigenstate of the Hadamard gate. $\ket{T}$ is an eigenstate of what $\cite{bk04}$ refers to as the $T$ gate, but to avoid confusion with the $\pi/8$-gate which also often carries that name, we will refer to as the $M_3$ gate as coined by Gottesman in \cite{g97}. It is defined by how it acts on Pauli matrices by conjugation:
\begin{equation} M_3^\dagger X M_3 = Y; \hspace{2mm} M_3^\dagger Y M_3 = Z; \hspace{2mm} M_3^\dagger Z M_3 = X. \label{eq:m3gate} \end{equation}

    Clifford eigenstates like $\ket{T}$ are valuable because \textit{any} qubit state can be projected onto an axis given by $\rho(r) = (1-|r|)I/2 + r\ket{T}\bra{T} $ with $-1 \leq r \leq 1$. This is achieved via an operation called \textit{twirling} \cite{cab12}:
$$ \rho \to \frac{1}{3}\rho + \frac{1}{3} M_3^\dagger \rho M_3 +  \frac{1}{3} (M_3^\dagger)^2 \rho M_3^2.$$

Twirling is achieved using classical randomness and Clifford operations. Now we can restrict our attention to density matrices whose Bloch vectors $\vec a$ are proportional to those of the $\ket{H}$ and $\ket{T}$ states.  In fact, since tight magic state distillation for $\ket{H}$ states is known to be possible \cite{R04}, we focus only on the $\ket{T}$ state. In this study we consider a class of input states where $r = |\vec a| \geq 0$:

$$\rho(r) = (1-r)\frac{I}{2} + r \ket{T}\bra{T} = \frac{I}{2} + \frac{r}{\sqrt{3}} \frac{X + Y + Z}{2} .$$

The derivation can also be done for arbitrary qubit mixed states. However, in this study we only consider the states on the positive half of the $\ket{T}$-axis for simplicity.

Now we are ready to investigate $\mathcal{R}(\rho^{\otimes n})$. What follows is an expansion of this expression from which the signed weight enumerator pops out naturally from how $\mathcal{R}$ acts on $\mathcal{P}_n$.

\clearpage
\begin{widetext}

    First we expand the tensor product $\rho^{\otimes n}$ using the Bloch sphere decomposition. In the expansion we encounter $\mathcal{P}^+_n$, the subset of $\mathcal{P}_n$ with $\lambda(P) = +1$.

$$\rho^{\otimes n} = \frac{1}{2^n}  \left(I + \frac{r}{\sqrt{3}} (X + Y + Z)\right)^{\otimes n} = \frac{1}{2^n} \sum_{P \in \mathcal{P}^+_n} \left(\frac{r}{\sqrt{3}}\right)^{\text{wt}(P)} P $$

    Now we apply $\mathcal{R}$, which acts linearly on the sum. All terms with $P \not\in G^\perp$ will have $\mathcal{R}(P) = 0$ and drop out. Therefore we can assume that $\mathcal{R}(P)= Q \in \{\pm I, \pm X, \pm Y, \pm Z\} $ for the remaining terms.  We also introduce the shorthand $\bar r = r / \sqrt{3}$.

    $$\mathcal{R}(\rho^{\otimes n}) =  \frac{1}{2^n} \sum_{P \in \mathcal{P}^+_n} \bar r^{\text{wt}(P)} \mathcal{R}(P) =  \frac{1}{2^n} \sum_{Q \in \{\pm I, \pm X, \pm Y, \pm Z\}} \sum_{\substack{P \in \mathcal{P}^+_n\\ \mathcal{R}(P) = Q }} \bar r^{\text{wt}(P)}    Q   $$

    Notice that for every $P \in G^\perp$ where $\mathcal{R}(P) = Q$, we also have $-P \in G^\perp$ where $\mathcal{R}(-P) = -Q$. Only one of $P, -P$ is in $\mathcal{P}^+_n$. Thus for each $P \in \mathcal{P}^+_n$, only one of the constraints $\mathcal{R}(P) = +Q$ and $\mathcal{R}(P) = -Q$ can be satisfied. We can eliminate this redundancy by switching to a sum over all $\mathcal{P}_n$ rather than just $\mathcal{P}_n^+$, and omitting the sum with $\mathcal{R}(P) = -Q$. \textit{It is crucial that we take sign into account by writing $\lambda(P) Q$ instead of $Q$:}
    When $\mathcal{R}(P) = Q$ and $\lambda(P) = 1$ then $P \in \mathcal{P}^+_n$ and $\mathcal{R}(P) = \lambda(P) Q = Q$ as before. When  $\mathcal{R}(P) = Q$ and $\lambda(P) = -1$, then  $-P \in \mathcal{P}^+_n$ so   $\mathcal{R}(-P) = - Q =  \lambda(P) Q $ correctly recovers the sum above.

    \begin{equation} \mathcal{R}(\rho^{\otimes n}) =  \frac{1}{2^n}  \sum_{Q \in \{I, X, Y, Z\}} \sum_{\substack{P \in \mathcal{P}_n\\ \mathcal{R}(P) = Q }} \bar r^{\text{wt}(P)} \lambda(P)Q \label{eq:esunsimp} \end{equation}

    \textbf{Definition 1:} \textit{For an $[[n,1]]$ stabilizer code $S$ with recovery map $\mathcal{R}$, the \textbf{signed quantum weight enumerator} for a Pauli operator $Q \in \mathcal{P}_1$ is: }

\begin{equation}
W_Q(\bar r) = \sum_{\substack{P \in \mathcal{P}_n\\ \mathcal{R}(P) = Q }}  \bar r^{\text{wt}(P)} \lambda(P) 
\end{equation}

Allowing the parameter $Q$ gives this definition the flexibility to describe coset weight enumerators, which is related to the weight distribution of translates in a classical linear code. Now we can simplify (\ref{eq:esunsimp}) and calculate the output state of the distillation procedure.

    $$  \mathcal{R}(\rho^{\otimes n}) = \frac{1}{2^n} \sum_{Q \in \{I, X, Y, Z\}}  W_Q(\bar r) Q =  \frac{1}{2^{n-1}} \left[ \frac{I}{2} W_I(\bar r) + \sum_{L \in \{X, Y, Z\}} \frac{L}{2} W_{L}(\bar r)  \right] $$

    \begin{equation}  \rho' = \frac{\mathcal{R}(\rho^{\otimes n})}{\text{Tr}( \mathcal{R}(\rho^{\otimes n}))} = \frac{2^{n-1}}{ W_I(\bar r)} \mathcal{R}(\rho^{\otimes n})  =  \frac{I}{2} + \sum_{L \in \{X, Y, Z\}} \frac{L}{2} \frac{W_{L}(\bar r) }{W_I(\bar r)}  \end{equation}

From this calculation we can draw two conclusions, which we package into a theorem.
\\

    \textbf{Theorem 1:}  Let $S$ be an $[[n,1]]$ stabilizer code with a recovery map $\mathcal{R}$. Let the input state $\rho$ have Bloch vector $\vec a = (r /\sqrt{3})(1,1,1)$ and let $\bar r = r/\sqrt{3}$. When using this code to distill magic states via the map $\rho \to \rho' = \mathcal{R}(\rho^{\otimes n})/\text{Tr}( \mathcal{R}(\rho^{\otimes n}))$,
 then:

\begin{enumerate}
    \item for operators and Bloch vector components $L \in \{X,Y,Z\}$, the output state $\rho'$ has a Bloch vector $\vec a'$ whose components are given by $a_{L}' = W_{L}(\bar r) /W_I(\bar r)$, and
    \item the procedure succeeds with probability $W_I(\bar r)/2^{n-1}$.
\end{enumerate}


\end{widetext}

\section{On Quantum Weight Enumerators}

Weight enumerators are key ideas in classical coding theory \cite{MWS}. In the theory of quantum stabilizer codes weight enumerators appear less frequently. In this section we speculate why this is the case, and discuss some previous work involving quantum weight enumerators. We also prove that the signed quantum weight enumerator is NP-hard to compute in general, but discuss why the proof may be unsatisfying for practical codes.

For quantum codes the situation is more complicated than that of classical codes. \textit{If the sign of operators is ignored}, then a quantum code corresponds to a classical code over $GF(4)^n$. Following the construction in \cite{CRSS96}, we define for $a \in \mathbb{F}_2^n$:
$$X(a) = \bigotimes_{i=0}^{n} X^{a_i}, \hspace{2mm} Z(a) = \bigotimes_{i=0}^{n} Z^{a_i}.$$

Then a signless Pauli operator is a pair of vectors $a,b \in \mathbb{F}_2^n$ and can be written as $P_{a,b} = X(a)Z(b)$. Pairs of vectors in $\mathbb{F}_2^n$ correspond to a vector space over $GF(4) = \{0, 1, \omega, \omega^2\}$ with $\omega+1 = \omega^2, \omega^3 = 1$, via the map $\phi(a, b) = a\omega + b\omega^2$. Using this connection to vector spaces we can define \textit{two} different inner products of Pauli operators:
$$P_{a,b} \cdot Q_{a,b} = a^T a + b^T b $$
$$P_{a,b} * Q_{a,b} = a^T b + b^T a $$

The standard inner product $\cdot$ is useful when viewing a code $G$ as a vector space and calculating the null space which we write as $G^{\perp}(\cdot)$. The \textit{symplectic} inner product $*$ is $0$ iff two operators commute, so $G^\perp(*)$ is the normalizer of $G$ (which we just called $G^\perp$ earlier). $G$ is abelian, so $G \subset G^\perp(*)$.

The most important fact about classical weight enumerators is the MacWilliams identity, which relates the weight enumerator of a code $C$ to its dual code $C^\perp$. But what is the dual of a quantum code, $G^{\perp}(\cdot)$ or $G^{\perp}(*)$? Since $G$ has dimension $n-1$ and is part of a space with dimension $2n$ (pairs of vectors of $\mathbb{F}_2^n$), both of these potential duals must have dimension $2n - (n-1) = n + 1$, which is too many to define a quantum code. Even if the dimension were correct, self orthogonality according to $*$ is not guaranteed. So the dual of a quantum code does not have a satisfying definition for $[[n, 1]]$ codes. It can only be done for $[[n,0]]$ codes, also known as stabilizer states. A quantum code's corresponding $GF(4)$ code has a dual, but that dual does not correspond to a quantum code. This limits the amount of meaning a quantum MacWilliams identity can have, possibly explaining why weight enumerators show up less in quantum coding theory.

However, that a quantum dual code cannot always be defined does not imply that a quantum MacWilliams identity cannot be useful. In 1996, Shor and Laflamme \cite{sl96} defined a weight distribution of a quantum code that equaled the weight distribution of the corresponding $GF(4)$ code. Since $G \subset G^\perp(*)$, the coefficient on $x^w$ in the distribution of $G$ must be less than the $x^w$ coefficient in the distribution of $G^\perp(*)$. This gives a set of inequalities that any code must satisfy. These inequalities were used to show the non-existence of degenerate 5-qubit codes and [[9, 1, 5]] codes. Quantum weight enumerators have also been used more recently: in 2017 Ashikhmin used the Shor-Laflamme weight enumerators to give fidelity lower bounds for stabilizer codes \cite{ashikhmin17}.

The Shor-Laflamme definition of quantum weight enumerators ignores the sign of the operators in $G$. Indeed, allowing negative weights complicates the mathematics, and disconnects the discussion from the well-studied classical codes over $GF(4)$. However, for magic state distillation sign plays a key role. For example, \cite{bk04} showed that the 5 qubit code generated by:
\begin{equation}S = \{XZZXI, IXZZX, XIXZZ, ZXIXZ\}\label{eq:5qubit}\end{equation}
is viable for distilling $\ket{T}$ states. But if the sign of any of these operators is flipped, the code becomes useless for this kind of distillation, so sign cannot be ignored. 

One connection to classical coding theory seems to remain: the NP-hardness of computing the weight enumerator. The difficulty of obtaining the weight distribution, or often even calculating the minimum weight of a code, plagues classical coding theory. Even when restricting to a particular family of codes, for many of the important families including BCH and Reed-M\"uller codes an efficient method for calculating the weight enumerator is not known \cite{MWS}. Since the signed quantum weight distribution $W_I$ characterizes magic state distillation, this hardness may explain why finding good codes for distillation is so challenging.

\textbf{Theorem 2:} Calculating $W_I$ for an arbitrary $[[n,1]]$ code $S$ is NP-hard.

\textit{Proof.} We give a standard reduction to the following NP-hard problem \cite{BMT78}: \textit{Given a set $S$ of $k$ binary vectors in $\mathbb{F}_2^n$ decide if the span of the vectors contains a vector with Hamming weight $w$.} We focus on the case when $k = n-1$. 

We construct a stabilizer code from the $n-1$ binary vectors, with a generator $X(a)$ for every $a \in S$. The stabilizer group $G$ of this quantum code contains $X(a)$ iff the span of $S$ contains $a$. This code has no negative weight operators, so no cancellation can occur in the weight enumerator. Thus if the $x^w$ coefficient in $W_I(x)$ is zero then $G$ contains no operators of weight $w$, so the span of $S$ cannot contain any vectors with Hamming weight $w$. Otherwise if the coefficient is nonzero such a vector must exist. \qed

In the general case there exist quantum codes with no negative phase operators, which is necessary for this proof to work. However, \cite{cb-bound} showed that under a reasonable set of assumptions every \textit{useful} $[[n,1]]$ code contains a negative operator. Proving hardness for these codes is less obvious because negative terms can cancel positive terms, so an $x^w$ coefficient of $W_I$ can be zero even though weight $w$ operators exist. However, brute-force searches show that this cancellation never happens for codes with $n \leq 5$: operators with different sign tend to have different weight.

If one can show that for these useful codes the reduction cannot hold, it would foster hope that a systematic way for constructing codes with good distillation thresholds exists. However, it may be more productive to simply find any non-trivial family of quantum codes for which the weight distribution can be calculated efficiently. Given our current understanding this is at least as hard as the classical version of the problem, which already seems very difficult.

\section{Analyzing distillation using weight enumerators}

In this section we apply signed quantum weight enumerators to the problem of finding the $\ket{T}$ state distillation threshold of a quantum code. For positive $r$, all stabilizer codes which map $r \to r'$ have a distillation threshold $r^*$, such that fidelity is only improved if $r$ is above the threshold. Otherwise fidelity is lowered: $r' < r$. The threshold $r^*$ is an unstable fixed point of the distillation. 

Often $r^* = 1$, meaning that the code is useless for $\ket{T}$-state distillation because it always reduces fidelity. The best known threshold is that of the 5 qubit code (\ref{eq:5qubit}) with $r^* = \sqrt{21}/7 \approx 0.655$. The lowest threshold imaginable is the surface of the stabilizer polytope $r^* = 1/\sqrt{3} \approx 0.577$, since if it were lower the code could distill states that can be prepared with Clifford operations. In 2010 \cite{cb-bound} showed that this threshold is unachievable by showing that the fidelity of the polytope surface state is decreased for all codes, so by continuity $r^*$ is strictly greater than $1/\sqrt{3}$.

This restriction shows that tight magic state distillation of $\ket{T}$ states is impossible using only a single code. This means that we must pursue infinite families of codes where $r^* \to 1/\sqrt{3}$ as $n \to \infty$ . In the previous section we argued how daunting this task is: we must give a family of codes where the weight enumerators are always known. In addition, we require that a property of the enumerators, $r^*$, approaches $1/\sqrt{3}$. The difficulty of the challenge might be partially circumvented if $r^*$ could be extracted from the code without knowing the weight enumerator, but no method for doing this is known.

Now we describe how to extract $r^*$ given enumerators $W_Q$ for all $Q \in \mathcal{P}_1$. Recall from theorem 1 that the output Bloch vector components are given by:
$$a_{L}' = \frac{W_{L}(\bar r)}{W_I(\bar r)}.$$
If we twirl this state back into the form $\vec a' = (r' /\sqrt{3})(1,1,1)$ then all components are transformed $ a_L' \to (a_X' + a_Y' + a_Z')/3$. This may first require a Clifford rotation of the output state into the positive octant, which just redefines $C_\text{decode}$ in $\mathcal{R}$. We now impose an additional requirement on the code $S$, which eliminates the need for twirling.

\textbf{Definition 2:} \textit{An $[[n, 1]]$ stabilizer code is \textbf{$\ket{T}$-axis preserving} if it has a recovery map $\mathcal{R}$ such that}
$$W_{X} = W_{Y}= W_{Z}.$$

Here it is necessary to mention $\mathcal{R}$ because the recovery map has the symmetry $\mathcal{R}(A) \to C\mathcal{R}(A)C^\dagger$ for qubit Clifford gates $C$. This gate can permute the set $\{\pm X, \pm Y, \pm Z\}$, altering the equation in the definition.

We do not consider a more general construction where the averaging over the different axes is taken into account, since this complicates matters and tends to worsen the performance of the code for $\ket{T}$ state distillation. Furthermore, there are so many quantum codes that restricting to codes that are easier to analyze is often a good idea. For a $\ket{T}$-axis preserving code, we can label all weight enumerators $W_{X},W_{Y},W_{Z}$ as just $W_{L}$: 

$$\frac{r'}{\sqrt{3}} = \bar r' = \frac{W_{L}(\bar r) }{W_I(\bar r)}.$$
Now we rearrange this equation to define a polynomial that has a root whenever distillation has a fixed point.

\begin{figure}[b]
    \includegraphics[width=0.51\textwidth]{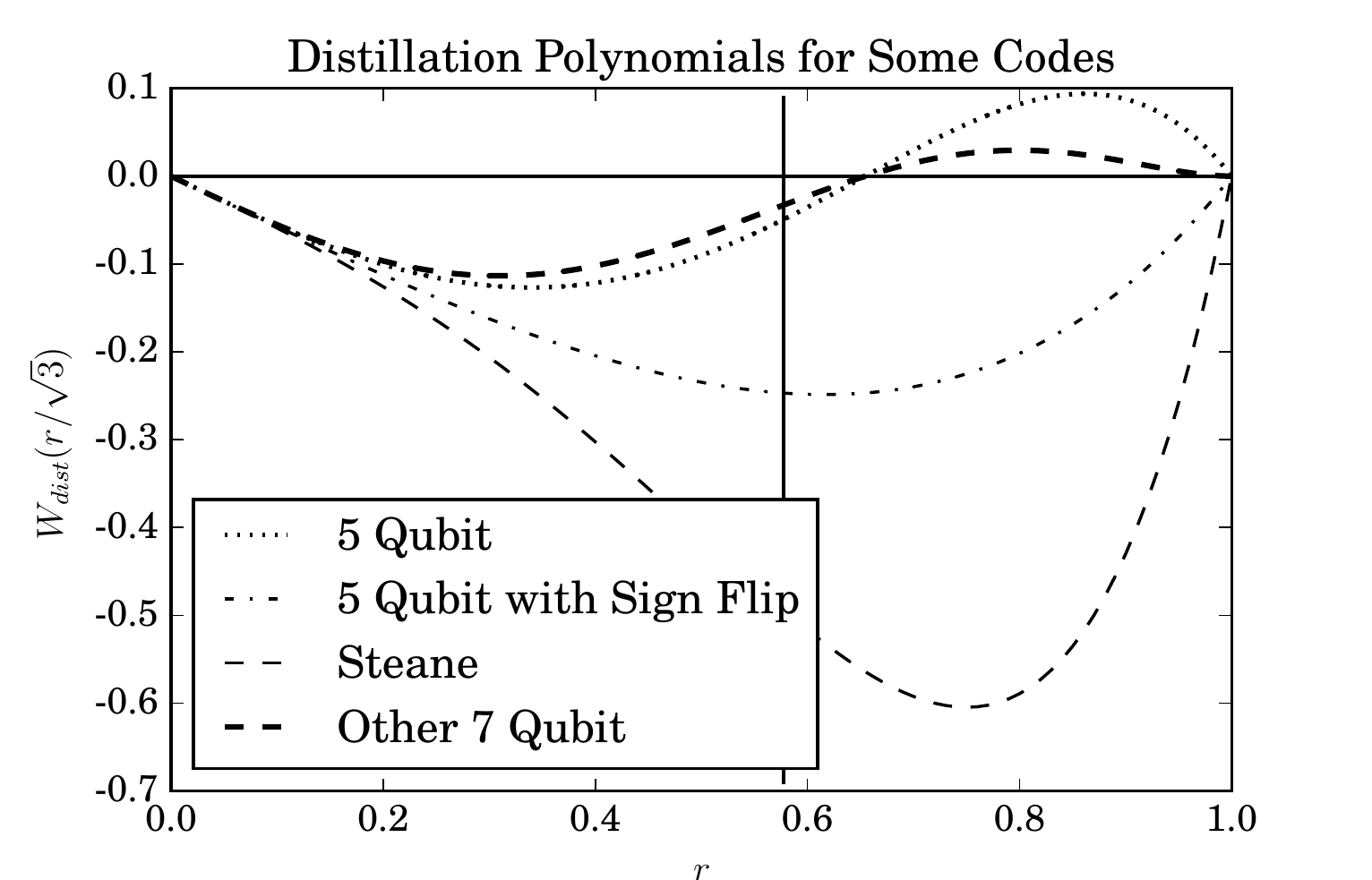}
    \caption{\label{fig:examples} Sketch of $W_\text{dist}(\bar r)$ for some codes. The 5 qubit code's (\ref{eq:5qubit}) polynomial is positive for some values, meaning it improves the fidelity of the input state. The same is true for a 7 qubit $M_3$-code defined by (\ref{eq:7qubit}). Both intersect the $r$-axis at $r^* = \sqrt{21}/7$. If the sign of a single generator in the 5 qubit code is flipped, it fails to distill any states. The Steane code (\ref{eq:steane}) also fails at distilling $\ket{T}$ states. The vertical bar at $r = 1/\sqrt{3}$ is the boundary of the stabilizer polytope.}
\end{figure}

\textbf{Definition 3:} \textit{If an $[[n, 1]]$ stabilizer code is $\ket{T}$-axis preserving then its \textbf{distillation polynomial} is:}
$$W_\text{dist}(\bar r) = W_{L}(\bar r)  - \bar r W_I(\bar r).$$
The sign of $W_\text{dist}(\bar r)$ determines if $\mathcal{R}$ improves or worsens input state fidelity. The potential values for $\bar r^* = r^*/\sqrt{3}$ are given by the roots of the distillation polynomial. We can see $W_\text{dist}(\bar r) = (\bar r' - \bar r)W_I(\bar r)$ is the increase in $\bar r$ scaled by the success probability (without the $2^{n-1}$), so greater values indicate better distillation performance.

The polynomial has some trivial roots. Since $I \in G$ the coset $\bar LG$ cannot contain $I$, so $W_L(0) = 0$. Thus $W_\text{dist}(0) = 0$, implying that the maximally mixed state is a fixed point. The $\ket{T}$ state $\bar r = 1/\sqrt{3}$ is a fixed point, since $\mathcal{R}$ contains only projections which map pure states to pure states. The result of \cite{cb-bound} that $W_\text{dist}(\bar r) < 0$ for all $0 < r \leq 1/\sqrt{3}$ can now be regarded purely as a property of stabilizer codes, proved via results from magic state distillation.

A search for $n = 5$ shows that all codes that are $\ket{T}$-axis preserving and have non-trivial fixed points have:
$$W_{L}(\bar r)  = -6 \bar r^5 + 10 \bar r^3,$$
$$W_I(\bar r) = 15 \bar r^4 + 1,\hspace{5mm} r^* = \sqrt{21}/7.$$
All these codes have the same distillation properties as the 5 qubit code (\ref{eq:5qubit}). Flipping the sign of any generator in any of these codes yields a useless ($r^* = 1$), but $\ket{T}$-axis preserving code. The search shows that no 5 qubit $\ket{T}$-axis preserving code does better than $r^* = \sqrt{21}/7$.

The Steane code is a 7 qubit code with generators:
\begin{equation} \label{eq:steane}
\begin{split}
    S = \{&XXXXIII, XXIIXXI, XIXIXIX,\\ & ZZZZIII, ZZIIZZI, ZIZIZIZ\}.
\end{split}
\end{equation}
This code achieves the optimal threshold for distillation of $\ket{H}$ states, and is $\ket{T}$-axis preserving. However it has $r^* = 1$, so it is no good for $\ket{T}$ states. We plot the distillation polynomials for the codes we just described in Fig. \ref{fig:examples}.

\section{$M_3$-codes}

In the previous section we restricted our attention to $\ket{T}$-axis preserving codes so that we could define the distillation polynomial $W_\text{dist}(\bar r)$. However, when numerically searching over the $O(2^{n^2})$ stabilizer codes this restriction does not significantly reduce the amount of work. In this section we restrict our set of codes with the goal of improving the runtime of search algorithms.  

\textbf{Definition 4:} \textit{An $M_3$\textbf{-code} is an $[[n, 1]]$ stabilizer code where the transversal $M_3$ gate, $(M_3)^{\otimes n}$, is a logical operator.}

\textbf{Theorem 3:} $M_3$ gates and $M_3$-codes have the following properties.
\begin{enumerate}
    \item The 5 qubit (\ref{eq:5qubit}) and Steane codes (\ref{eq:steane}) are $M_3$-codes.
    \item If $P \in \mathcal{P}_n$, then $\text{wt}(P) = \text{wt}((M_3^\dagger)^{\otimes n} P (M_3)^{\otimes n})$, i.e. acting on an operator by conjugation preserves weight. (This is true for all local Clifford gates).
    \item If $P \in \mathcal{P}_n$, then $\lambda(P) = \lambda((M_3^\dagger)^{\otimes n} P (M_3)^{\otimes n})$. $M_3, M_3^\dagger$ and $I$ are unique among single-qubit Clifford gates in their operator sign preservation.
    \item $M_3$-codes are $\ket{T}$-axis preserving.
    \item Generators of an $M_3$-code have even weight.
    \item If the weight of an $M_3$-code generator $P\in S$ is divisible by 4, then $\lambda(P) = +1$. Otherwise $\lambda(P) = -1$.
    \item $M_3$-codes have an even number of generators, which can be seen as pairs of $P$ and $(M_3^\dagger)^{\otimes n} P (M_3)^{\otimes n}$. 
\end{enumerate}
\textit{Proof.} See the appendix. \qed

Searching over $M_3$-codes is made easier by their structure. Since generators come in pairs by part 8), the number of choices for Pauli operators is cut in half. Furthermore the sign of each generator is fixed by its weight, so searching over sign choices is no longer necessary. Since $M_3$-codes are $\ket{T}$-axis preserving by design, there is no need to verify this property. We use a recursive algorithm that tries to cut the amount of exponential searching as much as possible and aborts if an odd weight operator is encountered. We also use a standard form developed for graph states \cite{vdNDM03} adapted to stabilizer codes by \cite{cb-bound} to even further reduce the number of possible generators. The algorithm is detailed in the appendix.

Ultimately, the search still runs in $O(2^{n^2})$ but the removal of some $O(2^n)$ factors improves runtime. A Python implementation on an quad-core computer was able to search $n=7$ overnight. All codes with a useful distillation threshold have:
$$W_L(\bar r)  = 18 \bar r^7 - 36 \bar r^5 + 10 \bar r^3,$$
$$W_I(\bar r) = -45 \bar r^6 + 15 \bar r^4 - 3\bar r^2 + 1, \hspace{8mm}r^* = \sqrt{21}/7.$$

An example of such a code is:
\begin{equation} \label{eq:7qubit}
\begin{split}
    S = \{ & XXXXIII, -XXXXYIX, YIZXIXI,\\& ZZZZIII, -ZZZZXIZ, XIYZIZI\}
\end{split}
\end{equation}
This code's distillation polynomial is shown in Fig.~\ref{fig:examples}.

\section{Conclusions}

We showed that for distillation of $\ket{T}$ states the signed quantum weight enumerators $W_Q$ characterize the distillation output. Since calculating these weight enumerators is just as challenging as it is for classical codes, finding an infinite family of quantum codes that achieves tight magic state distillation will be challenging. It is exciting how this insight may play together with other approaches to studying QCSI resource states, such as hidden variable theories, and how this would impact the theory of quantum codes.

Signed quantum weight enumerators give a direct method of calculating the distillation properties which can be tractable for small enough codes. A search over $n = 5$ codes showed that all $\ket{T}$-axis preserving codes that were useful had the same properties as the 5 qubit code, with threshold $r^* = \sqrt{21}/7$. By requiring a transversal $M_3$ gate which gives the code lots of structure, we were able to search $n = 7$, finding more codes with the same threshold as the 5 qubit code.

The astronomical size of the search space for even small values of $n$ is still an obstacle for finding good codes for distillation. By searching different regions of the space, such as the $M_3$-codes in this study, we may eventually develop some intuition for devising methods for construction. We suspect that signed quantum weight enumerators will be a focal point for such systematic methods. 

\section{Acknowledgments}
The author acknowledges financial support from the UT Austin Physics department. Scott Aaronson at the UT Austin CS department gave valuable advice throughout the project. The author also thanks Dan Browne at University College London for some helpful conversations.

\section*{Appendix A: Proof of Theorem 3\\ about the $M_3$ gate and $M_3$-codes}
\begin{enumerate}
    \item Verifiable by brute-force calculation. The Steane code's generators (\ref{eq:steane}) are already in the standard form described in part 8).
    \item Single-qubit Clifford gates acting (by conjugation) on single-qubit Pauli operators cannot map $\pm X, \pm Y, \pm Z$ to $\pm I$ or vice versa.

    \item Sign preservation follows immediately from the definition of the $M_3$ gate (\ref{eq:m3gate}). Since single-qubit Clifford gates permute $\{\pm X, \pm Y, \pm Z\}$ a sign-preserving gate must permute $\{X, Y, Z\}$ and $\{-X,-Y,-Z\}$ separately. A gate that is not $M_3, M_3^\dagger$ or $I$ could hold one operator, say $Z$, fixed, and swap $X \leftrightarrow Y$. Then $\ket{0},\ket{1}$ are eigenstates of the gate, so the gate must be $Z$ or $P$. But $Z^\dagger XZ = -X$ and $P^\dagger Y P = -X$.  


    \item Let us choose $\mathcal{R}$ such that $P \in \{\bar X, \bar Y, \bar Z\}$ all have $\lambda(P) = +1$. Each logical operator in $\{\bar X, \bar Y, \bar Z\}$ or $\{-\bar X,-\bar Y,-\bar Z\}$ can be reached by another element of the same set by applying $(M_3)^{\otimes n}$ once or twice. Since acting with $(M_3)^{\otimes n}$ preserves weight and sign, $\lambda(P) x^{\text{wt}(P)}$ is preserved by the map and all same-sign cosets have the same weight enumerator. Thus $W_{X} = W_{Y} = W_{Z}$ so the code is $\ket{T}$-axis preserving.
\\
\\
        \textit{ For the next three proofs we define a short hand. For $P \in \mathcal{P}_k$ we write $P' = (M^\dagger_3)^{\otimes k}P(M_3)^{\otimes k}$ and $P'' = (M_3)^{\otimes k}P'(M^\dagger_3)^{\otimes k} $. $k$ is to be inferred from the presence of a subscript as either $n$ or 1.}

    \item Consider a generator $P = (P_1\otimes P_2 \otimes ... P_n) \in G$ for an $M_3$-code. Notice that $P_i$ and $P_i'$ always anticommute unless $P_i = P_i' = I$. In an $M_3$-code $P\in G$ implies $P' \in G$. We also require $P$ and $P'$ commute, so $P_i$ and $P_i'$ must anticommute for an even number of qubits. Since the operators anticommute for all non-trivial qubits, $\text{wt}(P)$ must be even.

    \item Consider some $P\in G$ and $P' \in G$ in an $M_3$-code. By group closure their product must also be in $G$. For single-qubit Pauli operators $P_i$ different from $I$ notice that $P_i P_i' = i P_i'' $. Therefore $P  P'  = i^{\text{wt}(P)} \lambda(P) P'' $. But since $P''$ is also in $G$, and $-I\not\in G$, we must have $\lambda(P) = 1 / i^{\text{wt}(P)}$ which is  $i^{\text{wt}(P)}$ since $\text{wt}(P)$ is even. 

    \item Let us call a set $\{P, P', P''\} $ a `cycle'. If we have two cycles $\{P, P', P''\} $  and $\{Q, Q', Q''\} $  that share one element, e.g. $P = Q'$, then $P' = Q''$ and $P'' = Q$ so the cycles are identical. Thus two different cycles must be disjoint. We can thus decompose $G$ into some number of cycles and the identity. We can also multiply cycles together, so the cycles sort-of form a group on their own, except that there are three different ways to multiply cycles: $\{PQ, P'Q', P''Q''\},$ $\{PQ', P'Q'', P''Q\},$ $\{PQ'', P'Q, P''Q'\}$.

        We can select some set of `generator cycles' that are defined to be independent of one another. Generators of the code must be independent, and each cycle $\{P, P', P''\} $ only contains two independent generators since $PP' = P''$. Thus we can construct a generating set for the code out of $P, P'$ for each generator cycle. Since this construction can be done for all $M_3$-codes, all $M_3$-codes have an even number of generators.

\end{enumerate}

\section*{Appendix B: Algorithms for \\ searching over all codes and $M_3$-codes }

Here we describe a method for searching over quantum codes, and adapt the method for $M_3$-codes using the results from theorem 3. Naively, an $[[n, 1]]$ code is a list of $n-1$ Pauli operators, which each have $n$ elements of $\{I,X,Y,Z\}$ as well as a sign. We need 2 bits per qubit Pauli operator and one sign bit, resulting in $(n-1)(2n+1) = 2n^2 -n -1$ bits worth of freedom in the code, so $2^{2n^2 -n -1}$ possibilities must be checked. Our goal is to decrease the number of bits as much as possible, which will require some amount of polynomial overhead which we neglect in our runtime analysis.

We would like to immediately force the generators to be in the standard form described in \cite{cb-bound}. For an arbitrary code, we can use the Gram-Schmidt process to diagonalize the $X$ part of the first $n-1$ qubits of each generator. This way, for the $i$'th generator only the $i$'th and last qubits can be $X$ or $Y$. In other words, the $i$'th generator of $S$ lies in the space spanned by:
$$R_i = \{Z_1, Z_2, ..., Z_n, X_i, X_n\}$$

Now we see each generator needs just $n+2$ bits plus a sign bit, instead of $2n$ bits plus a sign bit, bringing the total number of bits to $(n-1)(n+3) = n^2 +2n-3$.

We know that generators in $S$ must be independent and commute. Say we are given a partial generating set with $i-1$ generators, $S_{i-1}$, and would like to construct a space containing all possible new generators. We can obtain this space by pruning $R_i$ so that all of the operators it spans commute with $S_{i-1}$, and it is disjoint with span($S_{i-1}$) excluding the identity.

A null space of a subspace according to an inner product $\diamond$ can be computed by truncating the space, by adding pairs of generators with inner product 1:

\begin{algorithm}
\caption{Null space calculation}
\begin{algorithmic}[1]
    \Procedure{Truncate($S$, $R$, $\diamond$)}{}
    \State $\textit{good} \gets \{\}.$
    \State $\textit{bad} \gets \{\}.$
    \For{$P \in S$}
        \For{$Q \in R$}
            \If{$P \diamond Q = 0$} $\textit{good} \gets \textit{good}\cup\{Q\}$
            \Else $\textit{ bad} \gets \textit{bad}\cup\{Q\}.$
            \EndIf
        \EndFor
        \State $R \gets \textit{good}$
        \For{$i \in [\text{len}(\textit{bad})-1]$} 
        \State $R \gets R\cup\{ (\textit{bad}[i]) (\textit{bad}[i+1])  \}$
        \EndFor
    \EndFor
    \State \Return{$R$}
\EndProcedure
\end{algorithmic}
\end{algorithm}

The space $\textsc{Truncate}(S_{i-1}, R_i, *)$ is the subspace of $R_i$ that commutes with $S_{i-1}$. We can also use the routine to calculate the null space $G^\perp(\cdot) = \textsc{Truncate}(S_{i-1}, \{Z_1,...,Z_n,X_1,...,X_n\},\cdot)$ which can be used as a parity check matrix to test for membership in $G = $ span($S_{i-1}$).

\begin{algorithm}
    \caption{Calculate if $P$ is in $G$ given $G^\perp(\cdot)$}
\begin{algorithmic}[1]
    \Procedure{inspace($P$, $G^\perp(\cdot)$)}{}
    \For{$Q \in G^\perp(\cdot)$}
        \If{$P \cdot Q = 1$} \Return \textsc{False} 
        \EndIf
    \EndFor
    \State \Return \textsc{True}
\EndProcedure
\end{algorithmic}
\end{algorithm}

We can use repeated membership tests to prune span($R_i$) to be disjoint from span($S_{i-1}$). We begin by removing any operators in $R_i$ contained in span($S_{i-1}$). This is not enough, because products of operators in $R_i$ could still be in span($S_{i-1}$). To deal with these, we move operators one-by-one into an output list $R_\text{out}$, and maintain a parity check matrix that checks for membership in span($S_{i-1} \cup R_\text{out}$).

\begin{algorithm}
    \caption{Truncate a basis $R$ to be disjoint from $S$}
\begin{algorithmic}[1]
    \Procedure{disjoint($S$, $R$)}{}
    \State $\textit{null} = \textsc{Truncate}(S, \{Z_1,...,Z_n,X_1,...,X_n\}, \cdot)$
    \State $R_\text{out} = \{\}$
    \While{$|R| > 0$}
        \State $R = \{r \in R \text{ if not } \textsc{inspace}(r, \textit{null})\}$
        \If{$|R| = 0$} break
        \EndIf
        \State $\textit{null} = \textsc{Truncate}(\{R[0]\}, \textit{null}, \cdot)$
        \State $R_\text{out} = R_\text{out} \cup \{ R[0] \}$
        \State $R = R / \{ R[0] \}$
    \EndWhile
    \State \Return $R_\text{out}$
\EndProcedure
\end{algorithmic}
\end{algorithm}

Now we have developed techniques to enforce three different constraints on the space of $i$'th operators to extend $S_{i-1}$: it must be in $R_i$, it must commute with $S_{i-1}$ and it must be disjoint (excluding the identity) from $S_{i-1}$. We employ a recursion technique to iterate over all quantum codes over $n$ qubits that satisfy these constraints. 

We also demand that for all operators $p$ we have $\text{wt}(p) \geq 2$ since smaller weights give trivial stabilizers.  There are so many codes that the procedure can get quite memory intensive. We therefore \textsc{analyze} the codes immediately and \textsc{aggregate} the results as they are computed. 

\begin{algorithm}
\caption{General code search}\label{alg:gensearch}
\begin{algorithmic}[1]
\Procedure{Search($n$, $S$)}{}
    \If {$|S| = n-1$} \Return $\textsc{analyze}(S)$ \EndIf
    \State $i \gets |S| + 1$.
    \State $R \gets \{Z_1, Z_2, ..., Z_n, X_i, X_n\}$. 
    \State $R \gets \textsc{disjoint}(S, R)$ 
    \State $R \gets \textsc{truncate}(S, R, *)$
    \State $\textit{results} \gets \{\}$
    \For{$P \in \text{span}(R)$}
        \If{$\text{wt}(P) \leq 1$} continue
        \EndIf
        \State $\textit{results} \gets \textit{results }\cup\textsc{search}(n, S\cup\{P\}) $
    \EndFor
    \State \Return $\textsc{aggregate}(\textit{results})$
\EndProcedure
\end{algorithmic}
\end{algorithm}

When diagonalizing the $X$ part of the generators we reduced the number of bits to search over per operator from $2n+1$ to $n+3$. Analyzing the number of bits needed after applying the $\textsc{disjoint}$ and $\textsc{truncate}$ routines, as well as requiring wt$(p) \geq 2$, is difficult because these steps do not reduce the number of generators consistently, especially if used together. However, they can only serve to reduce the total number of bits needed. Table 1 shows the maximum number of bits needed until a code was found, showing a performance improvement up to a factor of $2^{28-22} = 64$ for $n = 5$.

\begin{table}[h]
    \centering
\begin{tabular}{|c|c|c|c|c|}
    \hline
$n$ & 2 & 3 & 4 & 5 \\ \hline \hline
    $(n-1)(n+2)$ & 4 & 10 & 18 & 28 \\ \hline
Algorithm 4 & 4 & 9 & 15 & 22 \\ \hline
\end{tabular}

    \vspace{5mm}
    \caption{A quantum code in the standard form has $(n-1)(n+2)$ bits worth of freedom ignoring sign. By enforcing independence and commutativity the maximum number of bits actually needed by Algorithm 4 is less.}
    \vspace{-5mm}
\end{table}

The standard form described in \cite{cb-bound} is designed to eliminate degrees of freedom under which the code is invariant: we can multiply generators and rearrange qubits without changing the distillation properties of the code. However, the form does not completely eliminate all such redundancies. Consider a code whose generators have no $X$'s or $Y$'s. Then there is no $X$-part to diagonalize, and we are simply searching over all $n(n-1)$ binary matrices even though we would now be free to diagonalize the $Z$-part. Refining the standard form to account for this redundancy and possibly others could yield more efficiency boosts.

Now we discuss searching over $M_3$-codes, whose structure improves the runtime of the search algorithm. For general codes the \textsc{analyze} procedure must consider all possible signs of the generators, adding another $2^{n-1}$ possibilities. For an $M_3$-code the sign is determined by the weight of an operator by theorem 3 part 7), so this is not necessary. From theorem 3 part 8) we know that an $M_3$-code has a standard from $S = \{P, P', Q, Q', ...\}$, where $P' = (M_3^\dagger)^{\otimes n} P (M_3)^{\otimes n}$. Thus for every new generator $P$ we select, we can immediately add $P'$. This halves the number of recursion steps, and brings the total number of bits to $(n-1)(n+2)/2$. Doing so breaks the diagonalization of the $X$-part of the generators in $S$, but this is not an issue. We can simply restrict the standard form even further by first diagonalizing the $X$-part of the generators, then taking the first half of the operators $\{P, Q, ... \}$ and writing those as $\{P, P', Q, Q', ...\}$. 

\begin{algorithm}
\caption{$M_3$-code search}\label{alg:m3search}
\begin{algorithmic}[1]
\Procedure{M3search($n$, $S$)}{}
    \If {$|S| = n-1$} \Return $\textsc{analyze}(S)$ \EndIf
    \State $i \gets |S| + 1$.
    \State $R \gets \{Z_1, Z_2, ..., Z_n, X_i, X_n\}$. 
    \State $R \gets \textsc{disjoint}(S, R)$ 
    \State $R \gets \textsc{truncate}(S, R, *)$
    \State $\textit{results} \gets \{\}$
    \For{$P \in \text{span}(R)$}
        \If{$\text{wt}(P)$ is odd or $\text{wt}(P) = 0$} continue
        \EndIf
        \State $\textit{results} \gets \textit{results }\cup\textsc{search}(n, S\cup\{P, P'\}) $
    \EndFor
    \State \Return $\textsc{aggregate}(\textit{results})$
\EndProcedure
\end{algorithmic}
\end{algorithm}

The general search algorithm was able to search all $n = 5$ codes, and the $M_3$-code search algorithm was able to search all $n = 7$ codes. The design goal of these algorithms was to minimize the number of generators in $R$ before for loop on line 8 (for both algorithms) iterates over the exponentially large $\text{span}(R)$. However, for such small values of $n$, the exponential part of the algorithm may be comparable to the polynomial overhead incurred by truncating $R$. We therefore tested several versions of the algorithms that traded polynomial and exponential elements and found that for $n \leq 4$ most versions were quite similar in speed. For larger values of $n$ we found the full version of the algorithms were best.

Good search algorithms use standard forms of the output to reduce the size of the search space. One can look for general symmetries in all codes, or one can impose additional symmetry as in $M_3$-codes or codeword stabilized codes as in \cite{dh16}. One further symmetry present in all codes interesting for $\ket{T}$ state distillation is the distillation polynomial itself. Thus if a standard form can be developed using Clifford operations that preserve the distillation polynomial, the runtime can be improved even further.

Searching is most useful for finding small codes with good distillation properties in terms of threshold, distillation rate, and success probability. But for showing the possibility or impossibility of achieving tight distillation along the $\ket{T}$-axis there is no substitute for a systematic analysis.

\end{document}